\documentclass[prl,twocolumn,showpacs,superscriptaddress]{revtex4-2}
\usepackage{graphicx}
\usepackage{mathrsfs}
\usepackage{dcolumn}
\usepackage{bm}
\usepackage{amsmath}
\usepackage{amsfonts}
\usepackage{color}
\usepackage{makecell}
\usepackage{multirow}
\usepackage{booktabs}
\usepackage{amssymb}
\usepackage{hyperref}

\hypersetup{hypertex=true,
	colorlinks=true,
	linkcolor=blue,
	anchorcolor=blue,
	citecolor=blue}

\begin{document}

\title{Berry-phase-based Topological Charge in Quasicrystals and their Observable Features in Photonic System }

\author{Ziyi Chen}
\affiliation{Wuhan National High Magnetic Field Center $\&$ School of Physics, Huazhong University of Science and Technology, Wuhan 430074, China}
\author{Jinyu Zou}
\email[e-mail address: ]{jyzou@hust.edu.cn}
\affiliation{Wuhan National High Magnetic Field Center $\&$ School of Physics, Huazhong University of Science and Technology, Wuhan 430074, China}
\author{Jinhua Gao}
\affiliation{Wuhan National High Magnetic Field Center $\&$ School of Physics, Huazhong University of Science and Technology, Wuhan 430074, China}
\author{Gang Xu}
\email[e-mail address: ]{gangxu@hust.edu.cn}
\affiliation{Wuhan National High Magnetic Field Center $\&$ School of Physics, Huazhong University of Science and Technology, Wuhan 430074, China}
\affiliation{Institute for Quantum Science and Engineering, Huazhong University of Science and Technology, Wuhan, 430074, China}
\affiliation{Wuhan Institute of Quantum Technology, Wuhan, 430074, China}

\begin{abstract}

Topological charges based on Berry phase play the fundamental role in the topological physics. However, such topological charges remain unexplored in quasicrystals, impeding the systematic understanding of topological states in such quasiperiodic systems. In this work, by deriving all the allowed topological charges according to group representation theory and the corresponding low-energy effective Hamiltonians, we establish a universal framework for Berry-phase-based topological charges in two-dimensional quasicrystals. Taking the $C_{8v}$ quasicrystal as an example, we demonstrate and characterize a higher topological charge of $C=4$, which is inaccessible in conventional periodic systems. Applying our framework to photonic quasicrystals, we uncover that the circling of photon momentum around the charge gives a $C$ times winding of the electromagnetic field distribution pattern. Such observable feature provides a direct experimental method to probe the topological charges. Our work paves the way for exploring topological charges in quasiperiodic matter, and fundamentally bridges periodic and quasiperiodic topological band theories.

\end{abstract}

\maketitle

\renewcommand\thesection{\Roman{section}}
\section{INTRODUCTION}
Topological charge plays essential roles for the various topological states in condensed matter. Initially, by analogy with the Weyl fermions in elementary particles\cite{weyl1929electron}, Weyl points were discovered in Bloch bands\cite{wan2011topological,xu2011chern,huang2015weyl,lu2015experimental,soluyanov2015type,nagaosa2020transport}. These points are characterized by a topological charge ($|C| = 1$) encoding of the Berry phase, and trigger series of intriguing topological states and phenomena, such as the quantum Hall effect\cite{lonchakov2019observation}, Fermi arcs\cite{balents2011weyl}, chiral anomaly\cite{burkov2015chiral,ong2021experimental} and so on. Furthermore, high topological charge ($|C|\ge 2$) quasiparticles that go beyond the conventional Weyl points are discovered in crystal due to the protection of rotation symmetry\cite{fang2012multi,yang2014classification,bradlyn2016beyond,huang2016new,wang2018multiple,sinha2019transport,xue2022topological,arjas2024high}, which give rise to multiple chiral zeroth Landau levels and enhanced Berry curvature\cite{li2016weyl,xiong2022understanding,grushin2016inhomogeneous}, thereby resulting in nonlinear magnetotransport\cite{roy2022non,medel2024electric,focassio2024magnetic,burkov2017giant}, pronounced circular optical responses\cite{orenstein2013berry,menon2021chiral,saha2025unveiling}, and anisotropic optical conductivity characterized by distinct power-law\cite{ahn2017optical}. However, the topological charges are constrained up to $ \pm 3$ in two-dimension(2D) and $ \pm 4$ in three-dimension(3D) due to the limitation of the rotational symmetries in crystals\cite{schroter2020observation,fang2012multi,cui2021charge}, which motivates the pursuit of higher topological charge ($|C|> 4$) in systems beyond the crystals, such as quasicrystals.

Quasicrystals—materials exhibiting long-range orientational order while lacking translational symmetry—offer a natural platform with the rotational symmetries beyond crystals\cite{shechtman1984metallic,levine1984quasicrystals,levine1986quasicrystals,janot1994quasicrystals,steinhardt1987physics,walter2009crystallography,steurer2007photonic}. Although the absence of translational invariance invalidates the conventional Bloch band, effective band structures can be constructed via the theoretical frameworks of periodic approximation method\cite{goldman1993quasicrystals,wang2003photonic,yu2019dodecagonal,che2021polarization} and cut-and-project method\cite{jagannathan2021fibonacci,johnson2013computation,jiang2014numerical,he2024energy}, enabling the extension of topological charges to quasicrystalline systems. At present, however, discussions of topological charge in quasicrystals remain largely restricted to photonic systems, where the topological charge is defined through the winding of the polarization field around a polarization singularity\cite{zhen2014topological,zhang2018observation,ye2020singular,che2021polarization,arjas2024high}. Such definition is fundamentally different from the standard Berry-phase based one and therefore cannot serve as a foundation for a generalized topological physics of quasicrystals. This conceptual gap impedes systematic exploration of quasicrystal topological physics as in crystalline solids. Therefore, it is both timely and essential to directly extend the Berry-phase-based topological charges from crystalline systems to quasicrystals, especially the higher topological charges ($|C|> 4$).

In this work, applying the recently developed effective band method and group representation analysis to the 2D $C_{8v}$ quasicrystal, we demonstrate that a higher Berry-phase-based topological charge of $C = 4$, exceeding the limit of 2D crystals, can be realized. Extending this method to generic $C_{nv}$ quasicrystals, we establish a universal framework for topological charges in quasiperiodic systems. To this end, we first prove the inevitable existence of doubly degenerate states in $C_{nv}$ quasicrystals. Each degenerate state can be well described by a pair of conjugate angular momentum eigenstates, $\{j, -j\}$ with $j \le (n-1)/2$. Using this basis, we construct a $2\times2$ effective Hamiltonian and obtain the low-energy electronic structures, which are consistent with the quasicrystal effective band theory. The resulting eigenstates enable us to define the Berry phase and subsequently calculate the corresponding topological charges. Consequently, we establish a general relation among the $n$-fold rotational symmetry, angular momentum, and topological charges, identifying that the highest topological charges can reach up to $C = n/2$ or $(n - 1)/2$ for even and odd $n$, respectively. To bridge these theoretical results with experimental observations, we utilize the pseudospin texture as the fundamental signature of the topological charge. In particular, we uncover that such pseudospin texture can be directly depicted by the real-space electromagnetic field in photonic quasicrystals, which means that the real-space field distribution repeats itself exactly $C$ times when the photon momentum encircles the topological charge by $2\pi$. These results provide a universal method to construct higher topological charges and enable the direct experimental detection of such charges in quasicrystals.

\section{Berry Phase Based Topological Charge In Quasicrystals}
In this section, based on the effective band method of quasicrystal, we find that the $C_{8v}$ quasicrystal band structure hosts a series of doubly degenerate points, for which the topological charges can be defined through the Berry phase. Analysis shows that these degeneracies correspond to three 2D irreducible representations (irreps) of $C_{8v}$ with angular momenta $\pm 1$, $\pm 2$, and $\pm 3$, carrying topological charges of 2, 4, and -2, respectively, which can be further characterized by the pseudospin texture winding number. In subsection B, we generalize the Berry-phase-based topological charge to \(C_{nv}\) quasicrystals by constructing the general $2\times 2$ low-energy effective Hamiltonian, and summarize the correspondence among symmetry, angular momentum, and topological charge in Table.~\ref{table_1}.

\subsection{Higher topological charges in the $C_{8v}$ quasicrystals}
Although the momentum is not a strictly good quantum number in quasicrystals due to the absence of translational symmetry, the large-scale average structure still allows an effective momentum-space band description around the $\Gamma$ point of a pseudo-Brillouin zone \cite{Janssen2007,jiang2014numerical,wang2022effective,he2024energy}.
In general, For any potential function $V(\boldsymbol{r})$ possessing quasiperiodicity, one can always find a minimal number of basis vectors $\mathbf{G}_j$ to form the primitive reciprocal vectors, so that the the potential function can be expanded as \cite{Janssen2007}:
\begin{equation}
V(\mathbf{r})=\sum_{\mathbf{G}\in\mathcal{L}} \hat{V}(\mathbf{G})\, e^{i\mathbf{G}\cdot\mathbf{r}} ,
\end{equation}
with $\mathcal{L}$ defined as the set of reciprocal wave vectors $\mathbf{G}=\sum_i m_i  \mathbf{G}_i, m_{i}$ is any integer. Using this generalized set of reciprocal vectors, both the Hamiltonian and the wave functions can be expanded in momentum space, which in turn allows the effective band structure to be solved. 

\begin{figure*}[htbp]
	\centering
	\includegraphics[width=1\textwidth]{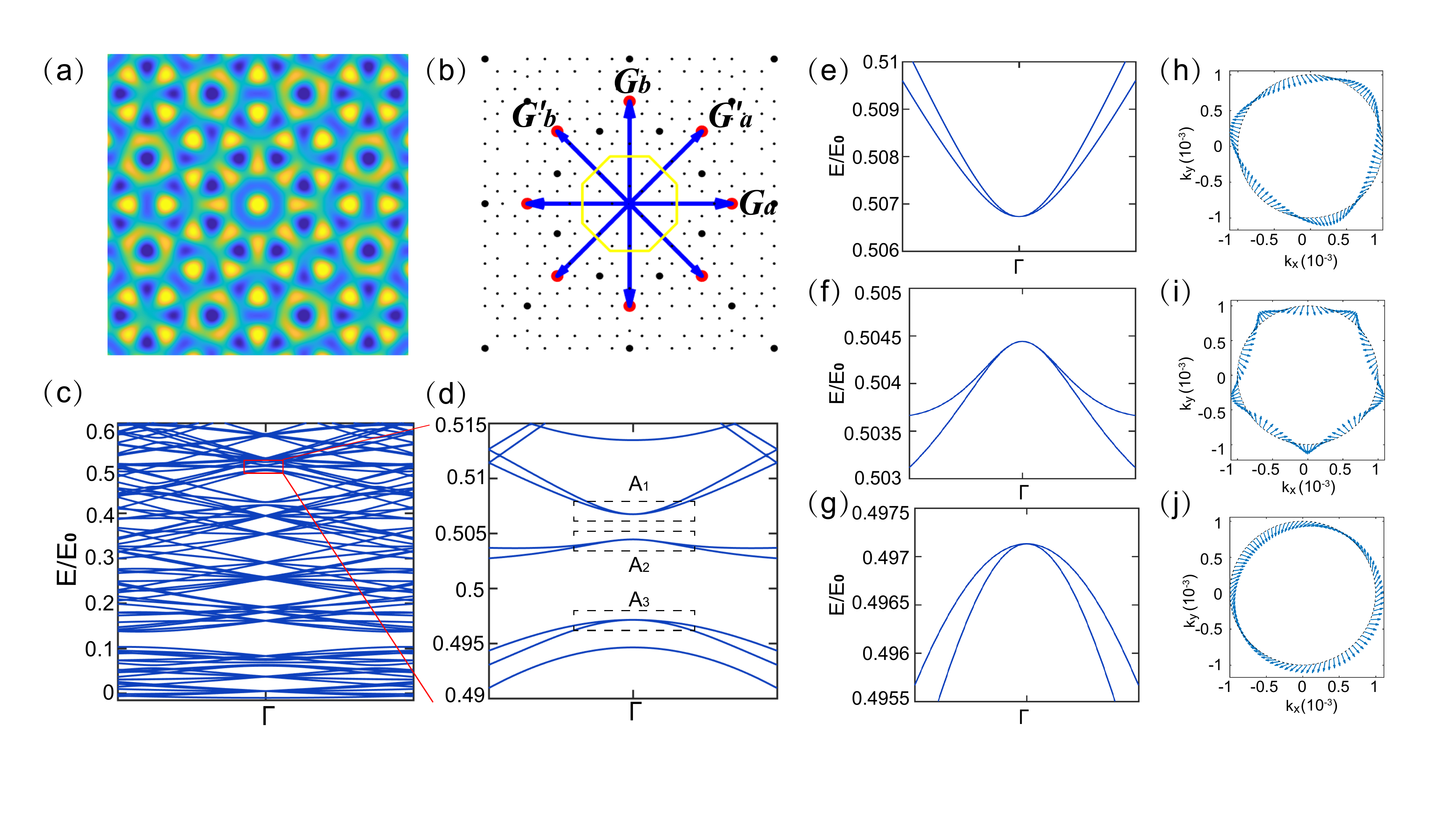}
	\caption{
		(a) Moiré quasicrystalline potential. 
		(b) Corresponding distribution of reciprocal-lattice points. 
		(c) Band structure after introducing a quasicrystalline potential $V$ near the $\Gamma$ point along the $\Gamma$–$\mathbf{G_a}$ direction, with the horizontal coordinate ranging from $\Gamma-0.1\mathbf{G_a}$ to $\Gamma+0.1\mathbf{G_a}$. Here $E_0=\hbar^2|\mathbf{G_a}|^2/2m$, $V=0.2E_0$, and $n_c=3$. The energy levels at $\Gamma$ point are grouped in eights.
		 (d) Zooms in on the bands framed by the red box in (c). The eight energy levels at $\Gamma$ point consist of three two-dimensional irreducible representations and two one-dimensional irreducible representations. (e)–(g) Zooms in on the three doubly degenerate points marked in (d). (h)–(j) Pseudospin textures on the lower bands in (e)-(g) with the momentum moving along a closed loop surrounding $\Gamma$ point. The winding numbers are 2, 4, -2, respectively.
	}
	\label{fig1}	
\end{figure*}

We apply the effective band method to a concrete quasicrystal system with the $C_{8v}$ symmetry, which describes a 2D electron gas in a quasiperiodical potential formed by two square periodic potentials twisted by $45^\circ$, as shown in Fig.~\ref{fig1}(a). Such a 2D quasicrystal has been experimentally realized in ultracold atom systems \cite{Viebahn2019}. We denote the primitive reciprocal vectors of the square potential as $\mathbf{G}_{a}$ and $\mathbf{G}_{b}$, and its rotated counterpart as $\mathbf{G}_{a}'$ and $\mathbf{G}_{b}'$. The four vectors $\{\mathbf{G}_{a},\mathbf{G}_{b},\mathbf{G}_{a}',\mathbf{G}_{b}'\}$ constitute the effective reciprocal vectors of the $C_{8v}$ quasicrystal \cite{jiang2014numerical}, as illustrated in Fig.~\ref{fig1}(b). Expanding the quasiperiodical potential by the effective reciprocal space, the Hamiltonian can be written as:
\begin{equation}
H=-\frac{\hbar^{2}}{2 m} \nabla^{2}+V \sum_{i} \cos \left( \mathbf{G}_{i} \cdot \boldsymbol{r}\right),
\end{equation}
with $ \mathbf{G}_{i} \in \{\mathbf{G}_{a},\mathbf{G}_{b},\mathbf{G}_{a}^{'},\mathbf{G}_{b}^{'}\} $.
To solve the Schrödinger equation $H\psi(\boldsymbol{r})=\varepsilon\psi(\boldsymbol{r})$, one can expand the wave function in plane-waves basis $\psi(\boldsymbol{r})=\sum_{\boldsymbol{q}} c_{\boldsymbol{q}} e^{i\boldsymbol{q}\cdot\boldsymbol{r}}$, where $c_{\boldsymbol{q}}$ is the complex coefficient and $\boldsymbol{q}$ denotes an effective momentum. Then the eigenvalue equation becomes
\begin{equation}\label{hq1}
\frac{\hbar^2 \boldsymbol{q}^2}{2m} c_{\boldsymbol{q}} +\frac{V}{2}\sum_{i} ( c_{\boldsymbol{q}-\mathbf{G}_{i}} + c_{\boldsymbol{q}+\mathbf{G}_{i}} )= \varepsilon c_{\boldsymbol{q}},
\end{equation}
which indicates that for a given $\boldsymbol{q}$, the potential $V$ couples the momenta set $\{\boldsymbol{k}| \boldsymbol{k}=\boldsymbol{q} + \mathbf{G}\} $, thereby defining an infinite-dimensional Hamiltonian matrix $H(\boldsymbol{q})$. To solve the energy spectrum, the reciprocal lattice vector $\mathbf{G}$ must be truncated so that $H(\boldsymbol{q})$ becomes a finite-dimensional matrix.
To preserve the $C_{8v}$ symmetry, we impose a truncation by taking the maximal reciprocal lattice vectors as $ \mathbf{G}_{max}=\sum_i n_i \mathbf{G}_i, |n_{i}|= n_c $, where $n_c$ is an integer. Here we choose $n_c=3$, and the resulted band structures near the $\Gamma$ point is shown in Fig.~\ref{fig1}(c). It is worth noting that the bands are grouped in eights, consisting of two sigle-bands and three doubly degenerate bands. Fig.~\ref{fig1}(d) zooms in on one such group, revealing three doubly degeneracies at the $\Gamma$ point.

To understand the emergence of these doubly degenerate points and determine their topological charge, we start our analysis of the states at the $\Gamma$ point with $V=0$. These states arise from shifting the parabolic band $\hbar^2 \boldsymbol{q}^2/2m$ with all the generalized reciprocal lattice vectors, i.e., $\boldsymbol{q} \rightarrow \boldsymbol{q}+\mathbf{G}$. Therefore, when $V=0$, these states at $\Gamma$  point are at least eight-fold degenerate (except the lowest states with energy $0$), corresponding to the rotational symmetry related reciprocal lattice vectors. Then we can reorganize each eightfold-degenerate states according to the irreducible representations of $C_{8v}$, i.e., construct the eigen states of the rotational symmetry by the following definition
\begin{equation}\label{construct} 
| \phi_{j} \rangle = \frac{1}{\sqrt{8}} \sum_{l=0}^7 (\omega_j)^l |\hat{C}_8^l \tilde{\mathbf{G}}\rangle ,
\end{equation}
where $j=-3,-2,\cdots,3,4$ is the angular momentum number and $\omega_j = e^{-i\pi j/4}$
is the corresponding phase factor. $\hat{C}_8^l$ denotes the $l$-fold application of rotational operator $\hat{C}_8$. 
$|\tilde{\mathbf{G}}\rangle, |\hat{C}_8\tilde{\mathbf{G}}\rangle, \cdots$ are the eigenstates of the parabolic band $\hbar^2 \boldsymbol{q}^2/2m$ with $\boldsymbol{q}$ taking a set of $C_8$-related reciprocal lattice vectors $\boldsymbol{q} = \tilde{\mathbf{G}}, \hat{C}_8\tilde{\mathbf{G}}, \cdots$. 
Such a construction satisfies
$\hat{C}_8 | \phi_{j} \rangle = \omega_{j} | \phi_{j} \rangle$, 
forming bases for the irreps of $C_{8v}$.

When $V \neq 0$, the eightfold degenerate states are coupled and consequently split according to the irreps. As illustrated in Fig.~\ref{fig1}(d), this splitting yields two 1D irreps and three 2D irreps, which we denote as $A_1$, $A_2$, and $A_3$, corresponding to angular momentum spaces $\pm 1$, $\pm 2$, and $\pm 3$, respectively. For such degeneracies, a topological charge can be defined via the Berry phase accumulated along a closed loop encircling the degenerate $\Gamma$ point, which has been shown to be quantized when the in-plane mirror symmetry is preserved \cite{Zou2023}. To compute this phase, a $2\times 2$ low-energy effective Hamiltonian should be constructed. Symmetry constraints dictate that the off-diagonal terms take specific forms depending on the irrep: $q_{-}^{2}$ for $A_1$, $q_{+}^{4}+q_{-}^{4}$ for $A_2$, and $q_{+}^{2}$ for $A_3$ (see the third row of  Table~\ref{table_1} for details), where $q_{\pm}=q_{x}\pm i q_{y}$ is the momentum near the $\Gamma$ point. Consequently, integrating around a loop enclosing the $\Gamma$ point yields Berry phases of $2\pi$, $4\pi$, and $-2\pi$, which correspond to topological charges of $2$, $4$, and $-2$, respectively. Notably, the degeneracy carrying a topological charge of $4$ has no counterpart in ordinary 2D crystals. Moreover, the decomposition of the eightfold degeneracy in Eq. \ref{construct} enumerates all 2D irreps of $C_{8v}$, which means that the symmetry-allowed topological charges---namely $2$, $4$, and $-2$---are all realized in this quasicrystalline system.

The topological charge can also be conveniently characterized by visualizing the Berry phase through the winding number of the pseudospin texture\cite{jin2026experimental,park2011berry}. For a two-component state, the pseudospin texture is defined as \(\langle \psi | \pmb{\sigma} | \psi \rangle\), where \(\pmb{\sigma}\) denotes the Pauli matrices and \(|\psi\rangle\) is the eigenstate at momentum near the degenerate point. The winding number \(C\) of the pseudospin texture along a closed loop enclosing the degeneracy gives the Berry phase \(C\pi\), thereby providing an intuitive characterization of the topological charge. As shown in Fig.~\ref{fig1}(e--j), the pseudospin textures of the lower band for \(A_1\), \(A_2\), and \(A_3\) irreps exhibit winding numbers \(2\), \(4\), and \(-2\), respectively. Such pseudospin texture naturely provides an efficient experiment approach to detect these topological charge, such as polarization-dependent angle-resolved photoemission spectroscopy (polarization-dependent ARPES) \cite{hwang2011direct}.

\subsection{General theory for topological charge in $C_{nv}$ quasicrystals}
Above discussions about the degeneracies and the topological charges in $C_{8v}$ quasicrystal can be generalized to a $C_{nv}$ quasicrystal with an \(n\)-fold rotation symmetry \(\hat{C}_n\) and an in-plane mirror symmetry \(\hat{M}\). Without loss of generality, we set the mirror reflection is along x-direction. These symmetry operations satisfy the relations:

\begin{equation}\label{eq:cnv_relations}
\hat{C}_n^{\,n}=1,\qquad \hat{M}^{2}=1,\qquad \hat{C}_n\hat{M}=\hat{M}\hat{C}_n^{-1}.
\end{equation}
Suppose that \(\phi\) is an eigenstate of \(\hat{C}_n\) with an integer angular momentum $j$, i.e., $\hat{C}_n \phi = e^{-i 2\pi j/n}\phi$. Then one has
\begin{equation}\label{eq:mirror_partner}
\hat{C}_n \hat{M}\phi
=\hat{M}\hat{C}_n^{-1}\phi
= e^{-i 2\pi (-j)/n}\hat{M}\phi,
\end{equation}
indicating that $\hat{M}\phi$ is a state with angular momentum $-j$. Since the angular momentum $j$ and  $n+j$ are equivalent, the orthogonality between $j$ and $-j$ asks $j$ to be $1,\dots,\left\lfloor\frac{n-1}{2}\right\rfloor$ (see the 2nd column of Table~\ref{table_1}),  and thus the pair of states $\{j,-j\}$ forms a 2D irrep, resulting in symmetry-protected double degeneracy at the $\Gamma$ point.

\begin{table*}[htbp]
	\caption{For different $C_{nv}$ group, the symmetry allowed topological charges and the corresponding effective Hamiltonians $H_{eff}$ that constructed in the specific angular momenta basis $\{j,-j\}$. The symbol $\left\lfloor x \right\rfloor$ labels the floor function that takes the largest integer less than or equal to $x$.}
	\label{table_1}
	\centering
	\renewcommand\arraystretch{1.6}
	\begin{ruledtabular}
		\begin{tabular}{cp{3.2cm}<{\centering}cp{3.2cm}<{\centering}cp{4.2cm}<{\centering}cp{4.2cm}} 
			
			$n$ & $j$ & $H_{\text{eff}}$ & Topological charge ($C$)  \\
			\midrule
			\multirow{2}{*}{5} 
			& 1 & $(a+q_+ q_-) I+bq_-^2 \sigma_+ +h.c.$ & 2 \\
			& 2 & $(a+q_+ q_-) I+bq_+ \sigma_+ +h.c.$ & -1 \\
			\midrule
			\multirow{3}{*}{8}
			& 1 & $(a+q_+ q_-) I+bq_-^2 \sigma_+ +h.c.$  & 2 \\
			& 2 & $(a+q_+ q_- +q_+^2 q_-^2) I+(b_1 q_-^4+b_2 q_+^4) \sigma_+ +h.c.$ & 4 \\
			& 3 & $(a+q_+ q_-) I+bq_+^2 \sigma_+ +h.c.$ & -2 \\

			\midrule
			$n$ & $j=1, 2, \dots, \lfloor \frac{n-1}{2} \rfloor$ 
			& $aI+(b_1 q_-^{2j} + b_2 q_+^{n-2j}) \sigma_+ + h.c.$ 
			& $ 2j \quad or \quad 2j - n$ \\

		\end{tabular}
		
	\end{ruledtabular}
\end{table*}
To describe the degeneracy, the \(2\times2\) low-energy effective Hamiltonian around the degenerate point can be constructed in the subspace spanned by \(\{j,-j\}\), in which the symmetry operators are represented as
$D(\hat{C}_n)=\exp\!\left(-i\frac{2\pi j}{n}\sigma_z\right), D(\hat{M})=\sigma_x$. 
Imposing these symmetry constraints, the low-energy effective Hamiltonian takes the form as \cite{fang2012multi}:
\begin{equation}\label{kp}
H(q_+,q_-) = f(q_+,q_-) \sigma_{+} + f^*(q_+,q_-) \sigma_- + g(q_+,q_-) I.
\end{equation}
The $\sigma_z$ term is prohibited by mirror symmetry. $I$ denotes the $2\times 2$ identity matrix and $\sigma_{\pm}=\frac{1}{2}(\sigma_{x} \pm i\sigma_{y})$. $g(q_+,q_-)$ is constrained to take the form $g=\sum_{l\in \mathbb{N}} a_l q_+^l q_-^l$ where $a_l$ are arbitrary real parameters. $f(q_+,q_-)$ is a complex function and constrained to satisfy the condition:

\begin{equation}\label{constrain_f}
e^{-i\frac{4\pi j}{n}}f\left(q_+,q_-\right) = f\left( q_+ e^{2\pi i /n}, q_- e^{-2\pi i /n} \right).
\end{equation}
One can expand $f$ in powers of $q_+$ and $q_-$ as
$f(q_+,q_-) = \sum_{n_1,n_2} A_{n_1 n_2} q_+^{n_1} q_-^{n_2}$,
where the integers $n_1$ and $n_2$ are constrained by Eq.~\ref{constrain_f} as $n_2 - n_1 \equiv 2j \ (\text{mod } n)$.
Therefore, to the lowest order, the off-diagonal $f$ term for a given angular momentum $j$ is expressed as
\begin{equation}
f_j(q_+,q_-) \propto q_-^{2j} \quad \text{or} \quad q_+^{n-2j},
\end{equation}
depending on which exponent is smaller. Since the term proportional to the identity matrix does not affect the Berry phase, the topological charge is entirely determined by the winding number of $f(q_+,q_-)$, which is equivalent to its exponent. Consequently, a degenerate point formed by angular-momentum eigenstates $j$ and $-j$ is characterized by the topological number
\begin{equation}
C=
\begin{cases}
2j, & 2j \le n-2j,\\[4pt]
2j-n, & 2j>n-2j.
\end{cases}
\end{equation}

In Table~\ref{table_1}, we summarize the allowed topological charges and the corresponding effective Hamiltonians for general $C_{nv}$ point-group systems, using $C_{5v}$ and $C_{8v}$ as representative examples. These results demonstrate that higher topological charges are naturally accessible in such quasiperiodic systems. Specifically, the maximum allowable charge is strictly bounded by the $n$-fold rotational symmetry, reaching $C = n/2$ for even $n$, and $C = (n - 1)/2$ for odd $n$.

The theoretical framework can be readily generated to 3D quasicrystals, such as icosahedral quasicrystal Ho-Mg-Zn \cite{fisher1999magnetic} and Cd–Yb \cite{Tsai2000} with fivefold rotational symmetry, Ta–Te quasicrystal systems with twelvefold rotational symmetry \cite{conrad2002ta97te60}. The coexistence of these unconventional rotational symmetries with time-reversal and inversion symmetries can protect the fourfold degeneracies at the high symmetry point. Their 3D $Z_2$ topological charge $N$ is determined by the angular momentum $j$ as $N=2j$, therefore the higher topological charges are expected in these quasicrystals.

\section{Topoligical Charges in Photonic quasicrystal and the detecting scheme}
Artificial systems, such as photonic and phononic structures, provide promising platforms for realizing quasicrystals. In this section, we extend the theories of effective band structures and topological charges to photonic quasicrystals, to show the existence of higher topological charges and further propose an experimental scheme for their detection. 

\subsection{Topoligical charges in photonic quasicrystal}

We first extend the band theory of photonic crystals to the photonic quasicrystals. In any photonic materials, the electromagnetic wave should satisfy the Maxwell's equations, which yield the following eigenvalue equation \cite{joannopoulos1997photonic}:
\begin{equation}\label{ph1}
\nabla \times (\dfrac{1}{\epsilon(\boldsymbol{r})} \nabla \times \mathbf{H}(\boldsymbol{r}))=\dfrac{\omega^2}{c^2} \mathbf{H}(\boldsymbol{r}),
\end{equation}
where $c$ is the speed of light in vacuum, $\omega$ is the frequency of the photonic mode, and $\mathbf{H}(\boldsymbol{r})$ is the magnetic field at position $\boldsymbol{r}$. The  relative permittivity $\epsilon(\boldsymbol{r})$ is periodic in photonic crystals. The inverse permittivity can be expanded in reciprocal space as $\epsilon^{-1}(\boldsymbol{r}) = \sum_{\mathbf{G}} \kappa(\mathbf{G}) \exp(i\mathbf{G} \cdot \boldsymbol{r})$, where $\mathbf{G}$ denotes all reciprocal lattice vectors. Using a plane-wave basis expansion for the magnetic field: $\mathbf{H}(\boldsymbol{r}) = \sum_{\mathbf{q}} \mathbf{H}_\mathbf{q} e^{i\mathbf{q} \cdot \boldsymbol{r}}$
with $\mathbf{q}$ being an arbitrary wave vector, and substituting into Eq. (\ref{ph1}) yields:
\begin{equation} \label{ph2}
\sum_{\mathbf{G}} \kappa(\mathbf{G}) \mathbf{q} \times \left[(\mathbf{q}+\mathbf{G}) \times \mathbf{H}_{\mathbf{q}+\mathbf{G}}\right] =\dfrac{\omega^2}{c^2} \mathbf{H}_{\mathbf{q}}.
\end{equation}
It gives the $\mathbf{q}$ dependent Hamiltonian $\mathcal{H}(q)$ that couples the momenta set $\{\boldsymbol{k}| \boldsymbol{k}=\boldsymbol{q} + \mathbf{G}\} $. Solving the Hamiltonian gives the photonic bands $\omega_n(q)$ and the eigen states $|\psi_n(q)\rangle=(\cdots,\mathbf{H}_{n,q+G},\cdots)^T$, where $n$ is the band index. Thus, the real space distribution of a photonic mode $(q,\omega_n(q))$ is given by the inverse Fourier transformation, which takes the Bloch form: 
\begin{equation} \label{ph3}
\mathbf{H}_{n,\mathbf{q}}(\boldsymbol{r}) = \sum_{\mathbf{G}} \mathbf{H}_{n,\mathbf{q}+\mathbf{G}} e^{i(\mathbf{q}+\mathbf{G})\cdot\boldsymbol{r}}.
\end{equation}

The above analysis can be directly extended to photonic quasicrystals. In quasicrystals, the momentum space $q$ is spanned by a set of effective primitive reciprocal vectors $\{\mathbf{G}_1, \mathbf{G}_2, \ldots, \mathbf{G}_N \}$, which are defined such that the integer linear combinations $\mathbf{G}=\sum_{i=1}^{N} m_i \mathbf{G}_i$ ($m_i$ is any integer) generate the complete set of reciprocal lattice vectors compatible with the quasicrystal's rotational symmetry \cite{Janssen2007}. Consequently, the inverse permittivity admits a Fourier expansion over this reciprocal vector set: $\epsilon^{-1}(\boldsymbol{r}) = \sum_{\mathbf{G}} \kappa(\mathbf{G}) \exp(i\mathbf{G} \cdot \boldsymbol{r})$. Substituting this expansion into Eq. (\ref{ph1}) yields expressions formally identical to Eqs. (\ref{ph2}) and (\ref{ph3}), except that the summation over $\mathbf{G}$ now runs over the quasicrystalline reciprocal lattice vectors. Therefore, the effective band structure can be constructed for photonic quasicrystals, and the theory of topological charges established in Section II is applicable.

\begin{figure}[htbp]
	\centering
	\includegraphics[width=0.5\textwidth]{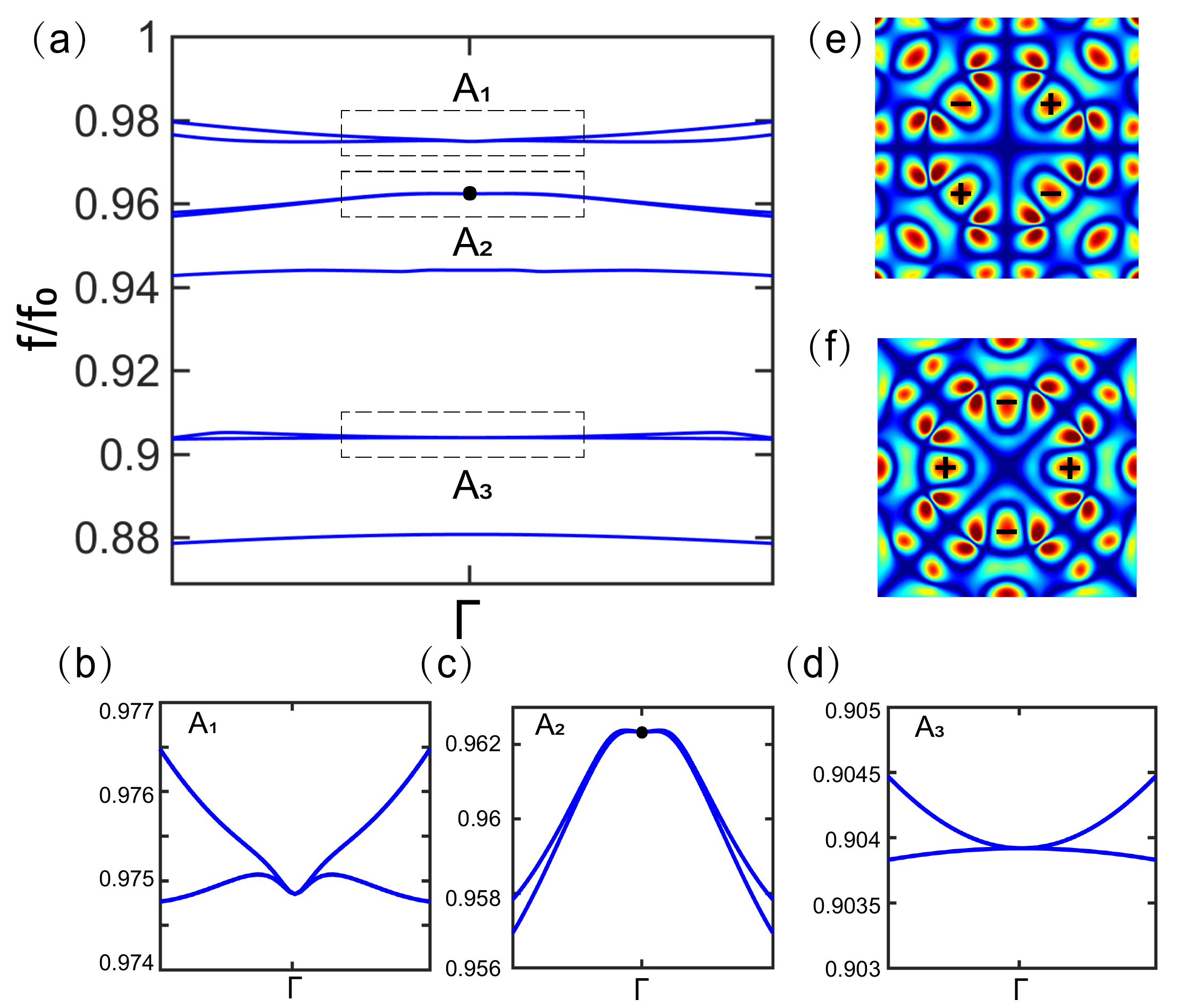}
	\caption{(a) Photonic band structure of the quasicrystal along the $\Gamma-G_a$ direction, with parameters set as $n_c=4$ and $f_0=c/2\pi a$, where $a$ represents the lattice corresponding to the primitive reciprocal vectors. (b)–(d) Zoomed-in views of the three doubly degenerate points marked in (a). (e)–(f) Two eigenmodes at the $A_{2}$ degeneracy point within a $5a \times 5a$ region, denoted as $d_{x^2-y^2}$ and $d_{xy}$, respectively.
	}

	\label{fig2}
\end{figure}
In the following, we consider a concrete example of photonic quasicrystal with $C_{8v}$ point-group symmetry. Without loss of generality, we focus on the TE mode, i.e., the magnetic field is along the out-of-plane $\mathbf{H}=(0,0,H_z)$. The photonic quasicrystal is constructed by modulating the permittivity to satisfy the $C_{8v}$ symmetry. The inverse permittivity is expressed as: $\epsilon^{-1}(\boldsymbol{r}) = 1+\alpha \sum_{i} \cos\!\left(\mathbf{G}_i \cdot \boldsymbol{r}\right)$, where $ \mathbf{G}_{i} \in \{\mathbf{G}_{a},\mathbf{G}_{b},\mathbf{G}_{a}^{'},\mathbf{G}_{b}^{'}\} $ are effective primitive reciprocal vectors, as illustrated in Fig.~\ref{fig1}(b). The modulation strength is set as $\alpha=0.25$. By numerically solving Eq.~(\ref{ph2}), one can obtain the band structure of the photonic quasicrystal. As we discussed in Section II.B, the band structure of a $C_{8v}$ preserved quasicrystal must be grouped in eights, consisting of two non-degenerate bands and three doubly degeneracies. Fig. \ref{fig2}(a) zooms in on such one group, exactly showing two single states and three doubly degenerate states (Fig. \ref{fig2}(b-d) respectively) at the $\Gamma$ point.

These doubly degeneracies correspond to the 2D irreps $A_1$, $A_2$ and $A_3$, which are formed in the angular momentum basis $\{j,-j\}=\{1,-1\},\{2,-2\},\{3,-3\}$ respectively. According to the classification list in Table~\ref{table_1}, they are characterized by the topological charges $C=2,4,-2$ respectively, revealing the realization of higher topological charge in the photonic quasicrystal.

\begin{figure}[htbp]
	\centering
	\includegraphics[width=0.5\textwidth]{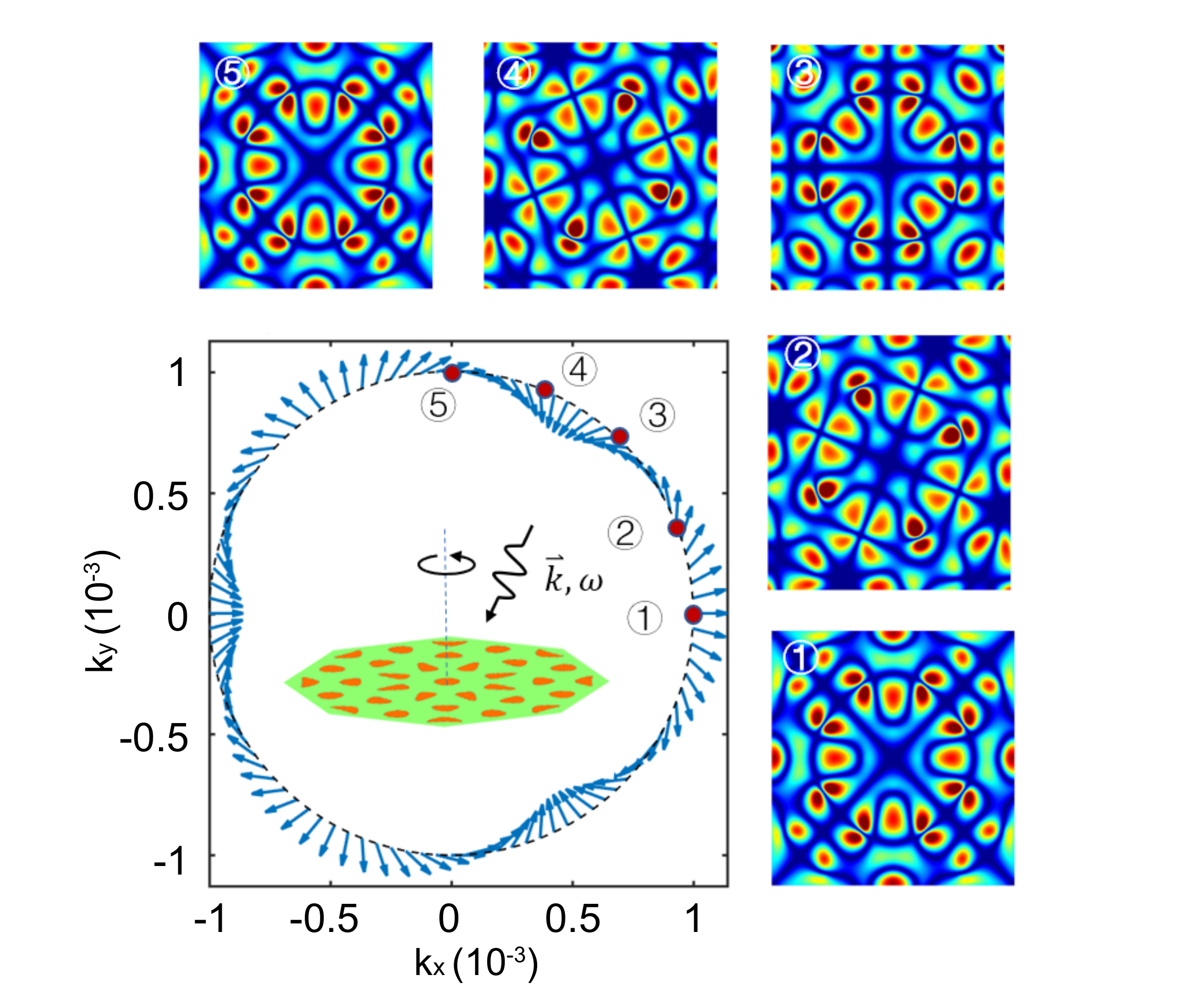}
	\caption{Observation of the $C=4$ topological charge. The central panel shows the momentum-space pseudospin texture and the experimental scanning scheme (inset). The corresponding real-space field patterns ($\textcircled{1}$–$\textcircled{5}$) repeat exactly after a $\pi/2$ rotation, directly demonstrating $C=4$.}
	\label{fig3}	
\end{figure}

\subsection{Detecting scheme of topological charges in photonic quasicrystal}
After obtaining the topological charges of the degenerate points, a key remaining question is how to experimentally detect these charges. As discussed in subsection II.B, the topological charge is determined by the number of times the pseudospin texture winds around the degeneracy point in momentum space. The key step is therefore to identify the physical observable corresponding to the pseudospin in photonic quasicrystals. Recall that the effective Hamiltonian near the degeneracy point is constructed in the eigenstate basis of rotational symmetry at the $\Gamma$ point, with angular momenta $j$ and $-j$. These correspond to photonic modes whose magnetic-field distributions satisfy $\hat{C}_n \mathbf{H}_{\mathbf{q}=0}^{\pm}(\mathbf{r}) = e^{\mp i2\pi j/n} \mathbf{H}_{\mathbf{q}=0}^{\pm}(\mathbf{r})$. These two high-symmetry modes correspond to the pseudospin up and down states and couple to each other when $\mathbf{q}$ deviates from the $\Gamma$ point. Their linear combination gives the photonic mode at momentum $\mathbf{q}$, $\mathbf{H}_{\mathbf{q}}(\mathbf{r}) = c_- \mathbf{H}_{\mathbf{q}=0}^{-}(\mathbf{r}) + c_+ \mathbf{H}_{\mathbf{q}=0}^{+}(\mathbf{r})$, where the coefficients $c_{\pm}$ are directly related to the pseudospin components as $(\langle \sigma_x \rangle, \langle \sigma_y \rangle)
= (c_-^*c_+, -i c_-^*c_+) + \text{h.c.}$. Consequently, the pseudospin reflects the magnetic-field distribution of the mode, which can be experimentally probed using techniques such as scanning near-field optical microscopy (SNOM) \cite{tsesses2025four}.

In this work, the two eigenstates belonging to the $A_2$ representation are labeled as $d_{x^2-y^2}$ (Fig. \ref{fig2}(e)) and $d_{xy}$(Fig. \ref{fig2}(f)). Under an eightfold rotation, they transform as:

\begin{equation} \label{ph4}
\hat{C}_8d_{x^2-y^2}=-d_{xy}, \hat{C}_8d_{xy}=d_{x^2-y^2}.
\end{equation}
By constructing $d_{\uparrow/\downarrow}=\frac{1}{\sqrt{2}} \left(d_{x^2-y^2} \pm id_{xy}\right)$, which satisfy $\hat{C}_8d_{\uparrow/\downarrow}=e^{\mp i \pi /4}d_{\uparrow/\downarrow}$, we obtain a pair of eigenstates with angular momentum quantum numbers $j=\pm 2$ and denote them as pseudospin up and down. Their linear combinations when $\mathbf{q}$ deviates from the $\Gamma$ point give the real-space field patterns, and directly reflect the pseudospin texture. As a concrete example, the mode displayed in Fig.~\ref{fig3}($\textcircled{1}$) exhibits a magnetic field pattern corresponding to $d_{xy}$. When projected onto the pseudospin basis $(d_\uparrow,d_\downarrow)$, this state is represented as $\frac{1}{\sqrt{2}} (1,1)$. From this we can compute the expectation values of the pseudospin operators, yielding $(\langle \sigma_{x} \rangle, \langle \sigma_{y} \rangle)=(1,0)$. In the lower-left panel of Fig.~\ref{fig3}, we choose five points $\textcircled{1} - \textcircled{5}$ in the loop circling the $\Gamma$ point, and present the magnetic field distribution respectively. It is obvious that the pattern repeats itself after a rotation of $\pi/2$ in momentum space, consistent with the winding number $C=4$ of the pseudospin texture. This result links the abstract topological charge to directly observable electromagnetic field distributions.

Based on the theoretical framework outlined above, we further propose a practical and feasible experimental detection scheme. The core of this scheme is to establish a direct correspondence between quasi-static scanning of the incident wavevector in momentum space and the observation of electromagnetic field distributions in real space. Specifically, the incident wavevector is scanned along a small closed path encircling the target degenerate point in momentum space, i.e., the incident angle of the light. Simultaneously, the corresponding distribution of localized electromagnetic fields on the sample surface is recorded. For a topological charge $C$, the real-space field pattern repeats itself after a rotation of $2\pi/C$ in momentum space. 
For example, a degenerate point with $C=4$ exhibits a real-space mode that repeats after a $\pi/2$ rotation in momentum space (Fig.~\ref{fig3}), whereas for $C=2$, a full rotation of $\pi$ is required. 
This provides a direct experimental method to verify the topological charges in quasicrystals.

It is worth noting that although our analysis is presented in the context of photonic quasicrystals, the underlying principles are general and can be extended to other artificial systems. For example, in phononic or acoustic quasicrystals, the pseudospin texture could be inferred by the evolution of real-space vibration or pressure-field patterns, which can be accessed experimentally by techniques such as scanning laser Doppler vibrometry or spatially resolved acoustic-field mapping \cite{kherraz2021experimental,yang2019observation}.

\section{CONCLUSION}

In summary, we establish a universal framework for Berry-phase-based topological charges in 2D $C_{nv}$ quasicrystals, and demonstrate that higher topological charges can be naturally accessible, with the highest charge reach up to $C = n/2$ or $(n - 1)/2$ for even or odd integer $n$ respectively. To experimentally observe these topological charges, we propose a feasible scheme in photonic quasicrystals: as the photon momentum circles the topological charge once, the electromagnetic field real-space distribution pattern will repeat exactly $C$ times. These results provide a foundational framework for extending Berry curvature and topological band theory from periodic crystals to quasiperiodic systems, and open avenues for investigating distinctive bulk–edge correspondence and novel transport phenomena, such as the  quantum hall effect in irrational magnetic field, enhanced circular dichroism, etc.

\section*{ACKNOWLEDGMENTS}
We would like to thank Bing Wang and Chengzhi Qin for helpful discussion about the photonic system. This work is supported by the National Natural Science Foundation of China (Grant No.~ 12404182, 12274154), and the National Key Research and Development Program of China (2024YFA1611200). The computation is completed in the HPC Platform of Huazhong University of Science and Technology.


%\bibliography{ref.bib}
\end{document}